\documentclass[12pt]{article}
\usepackage[cp1251]{inputenc}
\usepackage[english]{babel}
\usepackage[usenames]{color}
\usepackage{cite}

\def\pfrac#1#2{\frac{\partial #1}{\partial #2}}
\usepackage{amssymb,  amsfonts,amsmath}
\usepackage{hyperref}
\renewcommand{\dfrac}[2]{\frac{d#1}{d#2}}
\def\eqa#1{\begin{equation}\begin{aligned}#1\end{aligned}\end{equation}}
\def\eq#1{\begin{equation}#1\end{equation}}
\def\eqs#1{\begin{eqnarray}#1\end{eqnarray}}

\def\seq#1{\begin{equation*}#1\end{equation*}}
\def\seqs#1{\begin{equation*}\begin{split}#1\end{split}\end{equation*}}

\def\N{\mathbb{N}}
\def\Z{\mathbb{Z}}
\def\Im{\hbox{Im}}

\begin{document}

\begin{center}
{\Large  Generalized  symmetries and integrability conditions  for hyperbolic type semi-discrete equations}

\vskip 0.2cm

{Rustem  N. Garifullin}\footnote{e-mail:  rustem@matem.anrb.ru}\\

{Institute of Mathematics, Ufa Federal Research Centre, Russian Academy of Sciences,\\
112, Chernyshevsky Street, Ufa 450008, Russian Federation,\\
Bashkir State University, Validy str. 32, Ufa 450077, Russia, Russian Federation,}

\bigskip
{Ismagil T. Habibullin}\footnote{e-mail: habibullinismagil@gmail.com}\\

{Institute of Mathematics, Ufa Federal Research Centre, Russian Academy of Sciences,\\
112, Chernyshevsky Street, Ufa 450008, Russian Federation}\\
\bigskip


\end{center}
{\begin{flushright} {\it To the memory of A.B. Shabat and R.I. Yamilov}\end{flushright}}
\begin{abstract}

In the article differential-difference (semi-discrete) lattices of hyperbolic type are investigated from the  integrability viewpoint. More precisely we concentrate on a method for constructing generalized symmetries. This kind integrable lattices admit two hierarchies of generalized symmetries corresponding to the discrete and continuous independent variables $n$ and $x$. Symmetries corresponding to the direction of $n$ are constructed in a more or less standard way while when constructing symmetries of the other form we  meet a problem of solving a functional equation. We have shown that to handle with this equation one can effectively use the concept of characteristic Lie-Rinehart algebras of semi-discrete models. Based on this observation, we have proposed a classification method for integrable semi-discrete lattices. One of the interesting results of this work is a new example of an integrable equation, which is a semi-discrete analogue of the Tzizeica equation. Such examples were not previously known.

\end{abstract}
\bigskip

{\it Keywords:} generalized symmetry,  characteristic vector field, integrability
conditions, semi-discrete Tzizeica equation.

\section{Introduction}

We consider nonlinear differential-difference equations of the following form
\begin{equation}\label{dhyp}
u_{n+1,x}=f(u_{n,x},u_{n+1},  u_{n},x )
\end{equation}
where the sought function $u_n=u_n(x)$ depends on integer $n$ and real $x$ and $u_{n,x}$ denotes the derivative of $u_{n}(x)$ with respect to $x$. 
Below we use also expressions $u_{n,t}$ and $u_{n,\tau}$ denoting the derivatives of $u_n$ with respect to $t$ and $\tau$. 

Lattice \eqref{dhyp} can be regarded as a semi-discrete  version of the hyperbolic type partial differential equation. 
To preserve the parity of the forward and backward directions, we lay down the following condition. We assume that equation \eqref{dhyp} can be uniquely rewritten as
\begin{equation}\label{tdhyp}
u_{n-1,x}=\tilde f(u_{n,x},u_{n-1},u_{n},x).
\end{equation}

By analogy with the hyperbolic PDE we introduce two sets of dynamical variables for the lattice \eqref{dhyp} corresponding to two characteristic directions $x=const$ and $n=const$, first consists of the shifts of the variable $u_n$: $S_n:=\{u_k\}_{k=-\infty}^{+\infty}$ and the second contains the derivatives of $u_n(x)$ with respect to $x$ : $S_x:=\{\frac{d^k}{dx^k}u_n\}_{k=1}^{+\infty}$.
As usual, dynamical variables can be viewed as independent.  Unlike the  class of evolutionary type semi-discrete equations, such as the well-known  Volterra and Toda like chains, the class of lattices \eqref{dhyp} received less attention. However it should be noted that there are many examples of integrable lattices  from the class \eqref{dhyp} (see for instance, \cite{Yamilov1990}).  Equations of this type, often referred to as dressing chains \cite{vs93}, usually arise as iterated applications of the Backlund transformations for the nonlinear partial differential equations. In such a case the corresponding PDE is interpreted as a symmetry for the lattice, and then the Lax pair and hierarchy of the higher symmetries are readily found.  

Now we recall some necessary definitions (see cite ... for example). Let us given an evolutionary type PDE
\eq{u_{t}=g\left(x,u,u_{x},u_{xx},\ldots,\frac{\partial^Nu}{\partial x^N}\right)\label{kdvtI}}
of the order $N$. We assume that solution of the lattice \eqref{dhyp} depends on the variable $t$ and similarly solution of the PDE (\ref{kdvtI}) depends on $n$ such that the latter 
is converted into the form 
\eq{u_{n,t}=g\left(x,u_{n},u_{n,x},u_{n,xx},\ldots,\frac{\partial^Nu_n}{\partial x^N}\right).\label{kdvtn}}
Equation (\ref{kdvtI}) is called a symmetry of the lattice \eqref{dhyp} in the direction of $x$ if 
the flows defined by the equations \eqref{dhyp} and (\ref{kdvtn}) commute identically on the dynamical variables in $S_x\cup S_n$. In other words the following relation holds
\eq{ TD_xg-D_tf=0\quad \mod (1), (\ref{kdvtn})\label{symx}}
for all dynamical variables. Here $T$ is the shift operator of the discrete argument $n$ and $D_x$, $D_t$ -- operators of the total derivative with respect to the variables $x$ and $t$.

We call the evolutionary type lattice
\eq{u_{n,\tau}=h(u_{n-M},u_{n-M+1},\ldots,u_{n+M-1},u_{n+M},x)\label{voltI0}}
a symmetry of the lattice \eqref{dhyp} in the direction of $n$ if the flows defined by the equations \eqref{dhyp} and (\ref{voltI0}) commute with one another. More precisely we assume that equation 
\eq{ TD_xh-D_{\tau}f=0\quad \mod (1), (\ref{voltI})\label{symn}}
is satisfied identically on the set $S_x\cup S_n$. 

 The current state of symmetry approach in the integrability theory of the evolutionary type equations of the form \eqref{kdvtI} and \eqref{voltI0} can be found in the works \cite{msy87,y83,Yamilov06,ss82,ms12,is80,is79} and in the references therein.

Below in Section 2 we give a detailed explanation of the symmetries and illustrate the algorithm for finding them. Symmetries of the form (\ref{voltI0}) are found from an overdetermined system of differential equations, but the problem of constructing symmetries of the form (\ref{kdvtn}) is rather complicated and requires non-standard techniques using the concept of the characteristic operators of a semi-discrete chain. In Section 3 we discuss all the stages of the algorithm for constructing symmetries using the example of a semi-discrete version the  sine-Gordon equation. In section 4 we represent a new integrable semi-discrete model
\eq{u_{1,x}=u_{0,x}+(e^{-2u}+e^{-2u_1})+\sqrt{e^{2u}+e^{2u_1}}\label{tzi}}
that leads to the well-known Tzitzeika equation \cite{Tzitz,Dodd,Shabat,Mikh} in the continuum limit. Here we show that lattice (\ref{tzi}) does not admit any generalized symmetry of the order less than five in the direction of $x$. Its fifth order symmetry (see below equation (\ref{u_t}) with $\lambda_1=\lambda_2=1$) coincides with an integrable equation from the list in \cite{ms12}. The simplest generalized symmetry in the $n$-direction is a five-point evolutionary type lattice \eqref{u_tau}. To the best of our knowledge this equation is new. 

An amazing fact is that the lattice \eqref{tzi} does not define the Backlund transformation either for the Tzizeica equation or for its symmetry \eqref{91}, but defines the Backlund transformation for some third equation of the form (\ref{kdvtI}) connected with \eqref{91} by a very complex Miura-type transformation.

\section{Evaluation of the symmetries}

In this section, we discuss an algorithm for constructing generalized symmetries for the lattice \eqref{dhyp}. The problem is easily solved if a Lax pair or a lattice recursion operator is given. However, this method is not always available. For example, when solving the problem of integrable classification, we are dealing with equations of general form. It is curious that when we have only a lattice the problem of finding its generalized symmetry in the direction of $x$ may turn out to be nontrivial since we arrive at a nonlinear functional equation and probably this is one of the reasons why up to now  there is no any classification result for the soliton type lattices of the form \eqref{dhyp}.

Below we show how to reduce the over mentioned functional equation to a system of differential equations.

As it easily follows from the relations \eqref{symx} and \eqref{symn} the right hand sides $g$ and $h$ of the symmetries \eqref{kdvtI} and \eqref{voltI0} should satisfy the linearization of the equations \eqref{dhyp}
\eq{V_{n+1,x}=\pfrac{f}{u_{n,x}}V_{n,x}+\pfrac{f}{u_{n+1}}V_{n+1}+\pfrac{f}{u_{n}}V_n. \label{lineq}}

For the sake of convenience  we will use the following abbreviated notations: $u_k=u_{n+k}(x)$, $u_{k,1}=\dfrac{u_{n+k}}{x}$, $u_{k,2}=\dfrac{^2u_{n+k}}{x^2}$, \ldots. Since the lattices are autonomous with respect to $n$ these notations do not lead to confusion. Similarly we use expressions $u_{k,t}$, $u_{k,\tau}$ denoting the derivatives of the variable $u_{k}$ with respect to $t$ and $\tau$.

In the  abbreviated notations the symmetries (\ref{kdvtI}), (\ref{voltI0}) and the dynamical variables take the form
\eq{u_{0,t}=g(x,u_{0},u_{0,1},u_{0,2},\ldots,u_{0,N}),\label{kdvt}}
\eq{u_{0,\tau}=h(u_{-M},u_{-M+1},\ldots,u_{M-1},u_M),\label{volt}}
\eq{u_0,u_{1},u_{-1},u_{0,1},u_{2},u_{-2},u_{0,2},\ldots.\label{dynv}}
Obviously, all other variables $u_{k,m},\ m\in\N,k\in\Z-\{0\}$ can be expressed in terms of the dynamical variables \eqref{dynv} due to equation \eqref{dhyp}, its shifts and differential consequences. Completing the preparatory reasoning, we note that the equation \eqref{lineq} must be satisfied identically
for all dynamical variables \eqref{dynv} that are considered as independent ones.

We first explain how to find a symmetry in the  direction of $x$. To this end we substitute $V_n=g$, where $g$ is from \eqref{kdvt} into equation \eqref{lineq} and get a huge differential-difference functional equation, which is to be solved:
\eqa{\pfrac{}{x}g\left(x,u_{1},f,Df,\ldots,D^{N-1}f\right)+\sum_{i=0}^{N}D^{i}f\pfrac{}{u_{0,i}}g\left(x,u_{1},f,Df,\ldots,D^{N-1}f\right)=\\ \pfrac{f}{u_{n,x}}Dg(x,u,u_{0,1},u_{0,2},\ldots,u_{0,N})+\pfrac{f}{u_{n+1}}g\left(x,u_{1},f,Df,\ldots,D^{N-1}f\right)+\pfrac{f}{u_{n}}g\label{eq1}.}
Then we split down equation \eqref{eq1} with respect to the highest order derivative $u_{0,N+1}$. By collecting the coefficients in front of $u_{0,N+1}$ we get a relation :\eq{\pfrac{f}{u_{n,x}} T\pfrac{}{u_{0,N}}g(x,u_{0},u_{0,1},u_{0,2},\ldots,u_{0,N})=\pfrac{f}{u_{n,x}} \pfrac{}{u_{0,N}}g(x,u_{0},u_{0,1},u_{0,2},\ldots,u_{0,N})} 
that is simplified 
\eq{(T-1)\pfrac{}{u_{0,N}}g(x,u_{0},u_{0,1},u_{0,2},\ldots,u_{0,N})=0} 
and  easily solved
\eq{\pfrac{}{u_{0,N}}g(x,u_{0},u_{0,1},u_{0,2},\ldots,u_{0,N})=C(x),\quad C(x)\neq0.} 
Thus the sought function $g$ is partly specified and the symmetry \eqref{kdvt} gets the form:
\eq{u_{t}=C(x)u_{0,N}+g^{(1)}(x,u_{0},u_{0,1},u_{0,2},\ldots,u_{0,N-1}).\label{kdv1}} 
Now we turn back to equation \eqref{eq1} where instead of $g$ we substitute the rhs of the partly specified symmetry \eqref{kdv1}. The derivative of \eqref{eq1} with respect to $u_{0,N}$ after some simplification due to the relation, which can be easily proved by induction 
\eq{D^k f=\pfrac{f}{u_{0,1}}u_{0,k+1}+\left(kD\left(\pfrac {f}{u_{0,1}}\right)+\pfrac{f}{u_{0,1}}\pfrac{f}{u_1}+\pfrac{f}{u_0}\right)u_{0,k}+\ldots,k\geq 3,}
reads as:
\eq{ND\log \pfrac {f}{u_{0,1}}=(1-T)\pfrac{}{u_{0,N-1}}g^{(1)}(x,u,u_{0,1},u_{0,2},\ldots,u_{0,N-1})\label{int_cod1},\ N\geq3.} 

From \eqref{int_cod1} we obtain the first integrability condition for the lattice \eqref{dhyp}
\eq{D\log \pfrac {f}{u_{0,1}} \in \Im (1-T).} We stress that this condition doesn't depend on $N$.

Now we concentrate on (\ref{int_cod1}) that is in fact a functional equation, since it relates the values of the sought function $g$ at two different points $(x,u_{0},u_{0,1},u_{0,2},\ldots,u_{0,N-1})$ and $(x,u_{1}, u_{1,1}, u_{1,2}, \ldots,$ $u_{1,N-1})$. For the shortness we denote $z:=\pfrac{}{u_{0,N-1}}g^{(1)}(x,u_{0},u_{0,1},u_{0,2},\ldots,u_{0,N-1})$ and rewrite equation as follows
\eq{Tz-z=-ND\log \pfrac {f}{u_{0,1}}\label{z}.}

Functional equations of such kind usually arise when constructing symmetries of hyperbolic type equations on quadrilateral graphs. In our work \cite{GarifullinHabibullin} we proposed a method for solving them using the characteristic operators introduced earlier in \cite{h05}. Below, to solve equation \eqref{z}, we use the idea of \cite{GarifullinHabibullin}.

Since $z$ does not depend on the variable $u_1$ then by applying the operator $T^{-1}\pfrac{}{u_1}$ to both sides of \eqref{z} we reduce it to a differential equation:
\eq{Y_1z=-NT^{-1}\pfrac{}{u_1}(D\log \pfrac {f}{u_{0,1}}),\label{eq_l_1}} where the vector field $Y_1$ called the characteristic operator is given by \eq{Y_1=T^{-1}\pfrac{}{u_1}T=\pfrac{}{u_0}+\sum_{k=1}^\infty\left(T^{-1}\pfrac{}{u_1}D^kf\right)\pfrac{}{u_{0,k}}.}

In order to derive one more differential consequence of the functional  equation \eqref{z} we apply to both sides of that the operator $T^{-1}$  
\eq{z-T^{-1}z=-NT^{-1}D\log \pfrac {f}{u_{0,1}}\label{z-}}
and then apply to  \eqref{z-} the operator $T\pfrac{}{u_{-1}}$. As a result we get 
 \eq{Y_{-1}z=NY_{-1}(D\log \pfrac {f}{u_{0,1}})\label{eq_l_2}}
where $Y_{-1}$ is another characteristic operator defined as  \eq{Y_{-1}=T\pfrac{}{u_{-1}}T^{-1}=\pfrac{}{u_0}+\sum_{k=1}^\infty\left(T\pfrac{}{u_{-1}}D^k\tilde{f}\right)\pfrac{}{u_{0,k}}.} 
Characteristic operators $Y_1$, $Y_{-1}$ have been used earlier in \cite{Habibullin2008}, \cite{Habibullin2009} for the classification of the Darboux integrable lattices of the form \eqref{dhyp} in a particular case. 

Let's finalize the reasonings above. We have shown that function $z$ satisfies a system of two first-order linear equations (\ref{eq_l_1}, \ref{eq_l_2}) in partial derivatives with respect to the dynamical variables. However, $ z $ is a special solution, satisfying a very severe condition, indeed, by construction it does not depend on $ u_1 $ and $ u _ {- 1} $, despite the coefficients of the equations depend on these variables. To eliminate inappropriate solutions, we add to the system (\ref{eq_l_1}, \ref{eq_l_2}) the following two extra equations \eq{X_1z=\pfrac{}{u_1}z=0,\quad X_{-1}z=\pfrac{}{u_{-1}}z=0} 

Moreover, according to general theory of the systems of the first order linear equations in partial derivatives, z must satisfy, in addition to these four equations, several other equations derived from these equations by taking cross applications of the operators such as $[X_1,Y_1]z=X_1f^{(1)}$, $[Y_1,Y_{-1}]z=Y_1f^{(-1)}-Y_{-1}f^{(1)}$ etc., where the bracket stands for the commutator of the operators, $[Y,Z]=YZ-ZY$ (Jacobi's theorem).  In this way we arrive at a closed system of linear equations for partial derivatives $\pfrac{z}{u_{0,k}}$ \eq{\sum_{k=0}^{N-1}a_{s,k}\pfrac{z}{u_{0,k}}=b_s,\label{lin_sys}}
such that the further manipulations with the cross applications produce only equations that are linearly expressed through the earlier obtained equations.
Due to the well known Kronecker-Capelli theorem, system of linear equations \eqref{lin_sys} is compatible if and only if the rank of the coefficient matrix $A=(a_{s,i})$ is equal to that of the augmented matrix $B$ obtained from $A$ by adding the column of free terms $b_{s}$.  Thus the condition $rank(A)=rank(B)$ is necessary for existence of a symmetry of the form \eqref{kdvt}.

By solving equation \eqref{lin_sys} we find an explicit expression for the function\\ $z=\pfrac{}{u_{0,N-1}}g^{(1)}(x,u_{0},u_{0,1},u_{0,2},\ldots,u_{0,N-1})$ in terms of the rhs of the lattice \eqref{dhyp}. This allows one to get a further specification of the formula \eqref{kdv1}
\eq{u_{t}=u_{0,N}+p^{(1)}(x,u_{0},u_{0,1},u_{0,2},\ldots,u_{0,N-1})+ q^{(2)}(x,u_{0},u_{0,1},u_{0,2},\ldots,u_{0,N-2})\label{kdv2}}
for symmetry searched. Here $q^{(2)}$ is to be determined.

Applying several times the algorithm described above, we can completely determine the right side of the symmetry if it does exist. 
However in some cases we meet here additional problem related to the circumstance that equation \eqref{lin_sys} is a consequence of \eqref{z}, but these two equations are not equivalent. Hence \eqref{lin_sys} might have solutions which do not satisfy \eqref{z}. This problem is solved as follows. System \eqref{lin_sys} can be extended by adding some extra conditions obtained using the characteristic operators of a higher order (see \cite{Habibullin2008}, \cite{Habibullin2009}). 

By combining equation \eqref{z} and the following its consequence $T^2z-Tz=-NTD\log \pfrac {f}{u_{0,1}}$ we find 
\eq{T^2z=z-(1+T)ND\log \pfrac {f}{u_{0,1}}\label{z2}.}
Now we apply to the latter the operator $T^{-2}\pfrac{}{u_1}$ and since $\pfrac{z}{u_1}=0$ we obtain an equation
\eq{Y_2z=-T^{-2}\pfrac{}{u_1}(1+T)ND\log \pfrac {f}{u_{0,1}},\label{Y2z}} where $Y_2$ is a differential operator defined as 
\eq{Y_2=-T^{-2}\pfrac{}{u_1}T^2\label{Y2}.}

By applying similar manipulations we can derive equation 
\eq{Y_{-2}z=T^{2}\pfrac{}{u_{-1}}(T^{-1}+T^{-2})ND\log \pfrac {f}{u_{0,1}}\label{Y-2z},}
where
\eq{Y_{-2}=-T^{2}\pfrac{}{u_{-1}}T^{-2}\label{Y-2}.}

In the next step we derive linear equations similar to (\ref{Y2z}) and (\ref{Y-2z})
$$Y_3z=F^{(3)}\quad \mbox{and}\quad Y_{-3}z=F^{(-3)}$$
where $Y_3=-T^{-3}\pfrac{}{u_1}T^3$, $Y_{-3}=-T^{3}\pfrac{}{u_{-1}}T^{-3}$ and $F^{(\pm 3)}$ are some functions depending on $f$, its derivatives and shifts. One can proceed with such kind reasonings and derive more equations. However, as it is shown by examples to find the desired $z$ it is enough, as a rule, to use (\ref{Y2z}) and (\ref{Y-2z}), since the other equations do not produce additional conditions. 
Below in the next sections we illustrate  application of the algorithm with some examples. 

In the case of the lattice type symmetry \eqref{volt} the usual scheme can be used since the defining equation is an overdetermined differential (not functional!) equation. Indeed in this case linearized  equation \eqref{lineq} takes the form:
\eqa{D(h(u_{-M+1},u_{-M+2},\ldots,u_{M},u_{M+1},x))=\pfrac{f}{u_{n,x}}D(h(u_{-M},u_{-M+1},\ldots,u_{M-1},u_{M},x))\\+\pfrac{f}{u_{n+1}}h(u_{-M+1},u_{-M+2},\ldots,u_{M},u_{M+1},x)+\pfrac{f}{u_{n}}h(u_{-M},u_{-M+1},\ldots,u_{M-1},u_{M},x).\label{eq1_}}
At first we apply to both sides of the equation the operator $T^{-1}\pfrac{}{u_{M+1}}$ and get:
\eq{D\log\pfrac{h}{u_M}=(T^{-1}-T^{M-1})\pfrac f{u_1}\label{eqv}}
and then we apply to the same equation the operator  $\pfrac{}{u_{-M}}$ that gives:
\eq{D\log\pfrac{h}{u_{-M}}=(T-T^{-M+1})\pfrac {\tilde{f}}{u_{-1}}.\label{eqtv}}
The obtained relations \eqref{eqv} and \eqref{eqtv} are linear partial differential equations for unknowns $v=\log\pfrac{h}{u_M}$ and $\tilde v=\log\pfrac{h}{u_{-M}}$, respectively.  
First of them \eqref{eqv} reads:
\eq{\sum_{i=-M}^M D(u_i)\pfrac {v}{u_i}+\pfrac {v}{x}=(T^{-1}-T^{M-1})\pfrac f{u_1}.} Here coefficients depend not only on $u_i,-M\leq i\leq M$ but on $u_{0,1}$ as well. So we can add equations of the form: \eq{\label{eqva}\sum_{i=-M}^M \left(\pfrac{^j}{(u_{0,1})^j}D(u_i)\right)\pfrac {v}{u_i}=\pfrac{^j}{(u_{0,1})^j}(T^{-1}-T^{M-1})\pfrac f{u_1},j=1,...} and all their differential consequences.
From the system of equations \eqref{eqv}, \eqref{eqva} we can find function $v$ and then define the dependence of the function $h$ on $u_M$. In a similar way we can find dependence of $h$ on $u_{-M}$ from \eqref{eqtv} and its consequences.

Evidently equations \eqref{eqv} and \eqref{eqtv} imply the following necessary integrability conditions for the lattice \eqref{dhyp}:
\eq{(T^{-1}-T^{M-1})\pfrac f{u_1}\in \Im D,\quad (T-T^{-M+1})\pfrac {\tilde{f}}{u_{-1}}\in\Im D}

We note that defining equations having the form of a  functional equation similar to \eqref{eq1} often arise in the frame of the  symmetry approach.
Over the past two decades, a completely discrete analogue of the equation \eqref{dhyp}
\begin{equation}\label{ddhyp}
u_{n+1,m+1}=F(n,m,u_{n,m},u_{n+1,m},  u_{n,m+1})
\end{equation}
has been actively studied. The symmetry approach to this class of equations is developed in \cite{ly09,ly11,mwx11,GarifullinHabibullin}. Since for \eqref{ddhyp} both characteristic directions are discrete then the problem of constructing symmetries is related with functional equations. In our work \cite{GarifullinHabibullin} we suggested a method based on the characteristic operators, allowing to solve functional equation arising in the case \eqref{ddhyp}. In the recent article \cite{Xenitidis} by Xenitidis an alternative way to solve the functional equations for the symmetries of quad equations is developed.

\section{An illustrative example}

In this section we illustrate application of the algorithms outlined above by using the example of a semi-discrete version of the sine-Gordon equation \cite{b22}
\eq{u_{1,x}=u_{0,x}+\sin 2u_1+\sin 2u_0\label{eqsin}.}
For this lattice we derive generalized symmetries \eqref{kdv1} with $N=3$ and \eqref{volt} with $M=1$. 

To look for a generalized symmetry \eqref{kdv1} we use the defining equation \eqref{eq1} and its consequence \eqref{int_cod1} which for the lattice 
\eqref{eqsin} take, respectively,  the following form:
\eqa{(T-1)&(C'u_{0,3}+Cu_{0,4}+Dg^{(1)}(x,u,u_{0,1},u_{0,2}))\\ =&2\cos 2u_1(Cu_{1,3}+g^{(1)}(x,u_{1},f,Df))+2\cos 2u(Cu_{0,3}+g(x,u,u_{0,1},u_{0,2}))\label{eq1_sin}
} and \eqs{\label{eqg1_sin}
(1-T)\pfrac{}{u_{0,2}}g^{(1)}(x,u,u_{0,1},u_{0,2})=0.} 

Since equation \eqref{eqg1_sin} is easily solved explicitly, we do not need any characteristic operators at this stage. Indeed, obviously we have \eq{\pfrac{}{u_{0,2}}g^{(1)}(x,u,u_{0,1},u_{0,2})=C_1(x),}so formula \eqref{kdv2} reads as \eq{g(x,u,u_{0,1},u_{0,2},u_{0,3})=C(x)u_{0,3}+C_1(x)u_{0,2}+g^{(2)}(x,u,u_{0,1}).\label{sym}} Functions $C_1(x)$ and $g^{(2)}(x,u,u_{0,1})$ will be defined in the next step. At this  stage we  need in the operators $Y_{\pm1}$.
At first we evaluate the derivative of \eqref{eq1_sin} with respect to $u_{0,2}$:
\eq{(T-1)\pfrac{}{u_{0,1}}g^{(2)}(x,u,u_{0,1})=12C(u_{0,1}+\sin2u_1)(\sin 2u+\sin 2u_1)-2C'(\cos 2u+\cos 2u_1).\label{eqg2}}
Let us apply the operators  $T^{-1}\partial {}{u_1}$ and $Y_{-1}$ to both sides of \eqref{eqg2}. After some simplification we obtain two equations:
\eq{\pfrac{^2g^{(2)}}{u\partial u_{0,1}}=12C\sin 4u,\quad\pfrac{^2g}{u_{0,1}^2}=12Cu_{0,1}+2C'\tan 2u_0.} The consistency condition of these equations implies $C'(x)=0.$ Without loss of generality we can put $C(x)=1$ and define the following expression for the unknown function $q^{(2)}$ \eq{g^{(2)}=2u_{0,1}^3-(3\cos4u+C_2(x))u_{0,1}+g^{(3)}(x,u).} We specify the formula \eqref{sym} by virtue of the last relation and then substitute the found representation of the symmetry into equation \eqref{eq1_sin}. As a result we arrive at:
\eqa{\label{eq3_1}-4u_{0,1}^2C_1(\sin2u+\sin2u_1)+u_{0,1}\left((T-1)\pfrac{}{u_{0}}g^{(3)}+2C_1'(\cos2u+\cos2u_1)\right.\\-8\sin 2u_1C_1(\sin 2u+\sin 2u_1)\Big)+\ldots=0,} here the tail does not contain the variable $u_{0,1}$. By comparing the coefficients in front of the powers of the variable $u_{0,1}$  we find equations $$C_1(x)=0, \quad g^{(3)}=C_3(x)u_0+C_4(x)$$ that allow to specify equation \eqref{eq3_1}:
\seqs{(C_2'+C_3)(\sin 2u+\sin 2u_1)-2C_3(u \cos 2u+u_1\cos 2u_1)+C_3'(u_1-u_0)\\-2C_4(\cos 2u+\cos 2u_1)=0.}Since functions $u,\sin 2u,\cos 2u$ and $u\cos2u$ are linearly independent we get $$C_3(x)=0,\quad C_4(x)=0,\quad C_2(x)=c=\hbox{const}. $$

Thus, we have a complete solution of the defining equation \eqref{eq1} for the lattice \eqref{eqsin}
 \eq{g=u_{0,3}+2u_{0,1}^3-3u_{0,1}\cos4u -cu_{0,1}.} The last term in $g$  corresponds to the point symmetry of the lattice \eqref{eqsin} and we can take $c=0$. Finally, as expected, we obtain the well-known generalized symmetry of the lattice in the direction of  $ x $ :
\eq{u_{t}=u_{xxx}+2u_{x}^4-3u_x\cos4u .}

Now we turn to the symmetry in the direction of $ n $, which supposedly has the form \eq{u_{n,\tau}=h(u_{n-1},u_{n},u_{n+1}).\label{voltI}}  To look for the symmetry \eqref{voltI} we use the defining equation \eqref{eq1_} with $M=1$. In the case \eqref{eqsin} the differential consequences \eqref{eqv} and \eqref{eqtv} of the defining equation take the form
\seqs{(u_{0,1}-\sin 2u-\sin 2u_{-1})\pfrac{v}{u_{-1}}+u_{0,1}\pfrac{v}{u_0}+(u_{0,1}+\sin 2u_1+\sin 2u)\pfrac{v}{u_1}=2\cos2u-2\cos2u_1,\\(u_{0,1}-\sin 2u-\sin 2u_{-1})\pfrac{\tilde v}{u_{-1}}+u_{0,1}\pfrac{\tilde v}{u_0}+(u_{0,1}+\sin 2u_1+\sin 2u)\pfrac{\tilde v}{u_1}=2\cos2u_{-1}-2\cos2u.}
From these equations the functions $v=\log\pfrac{h}{u_1}$, $\tilde v=\log\pfrac{h}{u_{-1}}$  and $h$ are easily found  \eq{v=-2\log\cos(u_1-u_0)+\log C_1,\quad \tilde v=-2\log\cos(u_0-u_{-1})+\log C_2.} 
\eq{h=C_1\tan(u_1-u_0)+C_2\tan(u_0-u_{-1})+h^{(1)}(u_0).}
Now we get an equation for defining $h^{(1)}$, $C_1$ and $C_2$:
\eqa{(T-1)Dh^{(1)}(u)&-2(T+1)(h^{(1)}(u)\cos2u )\\&=2(C_2-C_1)\frac{(\cos2u_1-\cos2u)(\sin2u_1-\sin2u)}{\cos2u_1+\cos2u}.\label{eqh0}}
After differentiation of \eqref{eqh0} with respect to $u_{0,1}$ we get $$ (T-1)\pfrac{}{u_0}h^{(1)}=0.$$ Then we apply operator $T^{-1}\pfrac{}{u_1}-\pfrac{}{u_0}$ to \eqref{eqh0} and get a relation $C_2=C_1.$
Finally by applying the operator $\pfrac{}{u_0}$ to \eqref{eqh0} we obtain $$h^{(1)}(u_0)=0.$$
We set $C_1=1$, and finally find:
\eq{\partial_\tau u_0=\tan(u_{1}-u_0)+\tan(u_0-u_{-1})}
After the  M\"obius transformation of the form \eq{u_n=I\frac{v_n-1}{v_n+1},\  I^2=-1} we arrive at the well-known equation (V2) in the Yamilov list (see \cite{Yamilov06}, p. 596)
\eq{\partial_\tau v_0=-4v_0^2\left(\frac{1}{v_1+v_0}-\frac{1}{v+v_{-1}}\right).}

\section{A new example: integrable semi-discrete version of the Tzitzeika equation}

The celebrated Tzitzeika equation (see \cite{Tzitz,Dodd,Shabat,Mikh}) \seq{u_{xy}=e^u+e^{-2u}} is the most complicated example of the integrable equations of the Klein-Gordon type. The problem of integrable discretization of this equation has not been solved for a long time. Its completely discrete integrable versions, defined on a quadrilateral graph, were found relatively recently (see \cite{Adler,mx13,shl}).

It was generally believed that a semi-discrete analog does not exist at all, since Tzitzeika equation does not admit a first-order Backlund transformation. However the following lattice can be regarded as the desired semi-discrete version
\eq{u_{1,x}=u_{0,x}+\lambda_1(e^{-2u}+e^{-2u_1})+\lambda_2\sqrt{e^{2u}+e^{2u_1}}\label{tz}}
because it tends to Tzitzeika in the continuum limit, admits generalized symmetries on both directions of $x$ and $n$. Here $\lambda_1$ and $\lambda_2$ are arbitrary nonzero constants, which can be easily made equal to $1$. However we keep them for some computational reasons.
It can be shown that there is no any generalized symmetry of the form \eqref{kdv1} for $N=3$ as well as of the form \eqref{volt} with $M=1$ . We succeeded  to find generalized symmetries of the form \eqref{kdv1} with $N=5$: 
\eqa{\partial_t u_0=&u_{0,5}+5u_{0,3}(u_{0,2}-u_{0,1}^2-\lambda_2^2e^{2u}-\lambda_1^2e^{-4u})-5u_{0,2}^2u_{0,1}\\&-15u_{0,2}u_{0,1}(\lambda_2^2e^{2u}-4\lambda_1^2e^{-4u})+u_{0,1}^5-90\lambda_1^2u_{0,1}^3e^{-4u}+5u_{0,1}(\lambda_2^2e^{2u}+\lambda_1^2e^{-4u})^2,\label{u_t}}and of the form \eqref{volt} with $M=2$:
\eqa{\partial_\tau u=\Big((v^2-1)^2-4v_{-1}^2T^{-1}\Big)\frac{(v_{1}^2+1)(v_{-1}^2+1)}{(v^2(v_{-1}+1)^2+(v_{-1}-1)^2)(v_1(v+1)^2+(v-1)^2)},\\v=\sqrt{1+e^{2(u-u_1)}}+e^{u-u_1}.\label{u_tau}}
Equation \eqref{u_t} for fixed $n$ is known to be integrable and can be found in \cite[(3.8)]{ms12}. Equation \eqref{u_tau} and its modification \eqref{v_tau} seem to be new.

Let's get down to specific calculations. We look for the symmetry of the form \eqref{kdv1} with $N=5$ $$u_{0,t}=C(x)u_{0,5}+g(x,u,u_{0,1},u_{0,2},u_{0,3},u_{0,4}).$$  
Equations \eqref{eq1} and \eqref{int_cod1} take the form:
\eqa{(T-1)&(C'u_{0,5}+Cu_{0,6}+Dg^{(1)})\\ 
&=-2\lambda_1(T+1)e^{2u}(Cu_{0,5}+g^{(1)})
+\frac{\lambda_2}{\sqrt{e^{2u}+e^{2u_1}}}(T+1)e^{-2u}(Cu_{0,5}+g)\label{eq1_ts}
} and \eqs{\label{eqg1_ts}
(1-T)\pfrac{}{u_{0,2}}g^{(1)}(x,u,u_{0,1},u_{0,2},u_{0,3},u_{0,4})=5D\log \pfrac {f}{u_{0,1}}=0.} So,
according to the general scheme (see also the previously studied example) we obtain right away
\eq{g(x,u,u_{0,1},u_{0,2},u_{0,3},u_{0,4},u_{0,5})=C(x)u_{0,5}+C_1(x)u_{0,4}+g^{(2)}(x,u,u_{0,1},u_{0,2},u_{0,3}).}
We substitute the obtained representation of $g^{(1)}$ into equation \eqref{eq1_ts} and then differentiate the result  with respect to $u_{0,4}$. That gives rise to:
\eq{(T-1)\pfrac{}{u_{0,3}}g^{(2)}=5C(u_{0,1}L^2f+(f-u_{0,1})\dfrac{}{u_1}Lf)-C'Lf,\quad L=\dfrac{}{u_1}+\dfrac{}{u_0}.\label{eq1g2}}
By applying the operators  $T^{-1}\partial_{u_1}$ and $Y_{-1}$ to both sides of \eqref{eq1g2} we find, respectively:
\eqa{\sum_{k=0}^3Y_1(u_{0,k})\pfrac{^2}{u_{0,3}\partial u_{0,k}}g^{(2)}
=5C\left(8\lambda_1^2e^{-4u}-\lambda_2^2e^{-2u}-\frac{3\lambda_1\lambda_2+\lambda_2u_{0,1}e^{2u}}{\sqrt{e^{2u_{-1}}+e^{2u}}}+8\lambda_1u_{0,1}e^{-2u}\right)\\-C'\left(4\lambda_1e^{-2u}+\lambda_2\frac{e^{2u}}{\sqrt{e^{2u_{-1}}+e^{2u}}}\right)\label{eqq21}}
and
\eqa{\sum_{k=0}^3Y_{-1}(u_{0,k})\pfrac{^2}{u_{0,3}\partial u_{0,k}}g^{(2)}
=5C\left(\lambda_2^2e^{2u}-8\lambda_1^2e^{-4u}+\frac{3\lambda_1\lambda_2-\lambda_2u_{0,1}e^{2u}}{\sqrt{e^{2u_{1}}+e^{2u}}}+8\lambda_1u_{0,1}e^{-2u}\right)
\\-C'\left(4\lambda_1e^{-2u}+\lambda_2\frac{e^{2u}}{\sqrt{e^{2u_{1}}+e^{2u}}}\right).}
It can shown that these two equations are connected with each other by the transformation $\lambda_1\to-\lambda_1,\lambda_2\to-\lambda_2,u_{-1}\to u_1.$ Therefore, below we provide a detailed study only of the first of them.
We commute these equations with the operators $\pfrac{}{u_{-1}}$ and $\pfrac{}{u_{1}}$ respectively and obtain
\eqa{(u_{0,1}^2+u_{0,2}+\lambda_2^2e^{2u}\pm6\lambda_1u_{0,1}e^{-2u}&+7\lambda_1^2e^{-4u}) \pfrac{^2}{u_{0,3}^2}g^{(2)}+(u_{0,1}\mp 3\lambda_1 e^{-2u}) \pfrac{^2}{u_{0,3}\partial u_{0,2}}g^{(2)}\\ &+  \pfrac{^2}{u_{0,3}\partial u_{0,1}}g^{(2)}=-5C(u_{0,1}\pm3\lambda_1e^{-2u})-C'\label{eqq24}.}
The consistency condition of the system \eqref{eqq21}--\eqref{eqq24} implies $C'(x)=0$. So, without loss of generality we can put $$C(x)=1.$$ Solution of the system \eqref{eqq21}--\eqref{eqq24} is
\eq{g^{(2)}=-5(u_{0,1}^2-u_{0,2}+\lambda_1^2e^{-4u}+\lambda_2^2e^{2u})u_{0,3}+C_2(x)u_{0,3}+g^{(3)}(x,u,u_{0,1},u_{0,2}).}
Here $C_2(x)$ and $g^{(3)}(x,u,u_{0,1},u_{0,2})$ are functions to be defined in the next steps. Now we differentiate equation \eqref{eq1_ts} with respect to $u_{0,3}$ and get:
\seqs{(T-1)\pfrac{}{u_{0,2}}g^{(3)}=&-4C_1\Big(Au_{0,1}(\lambda_2+4\lambda_1A)+4A^2\lambda_1^2e^{-2u-4u_1}+\lambda_1\lambda_2(e^{-2u}+4e^{-2u_1})+\lambda_2^2e^{2u_1}\Big)\\&+10A^2\lambda_1(2u_{0,1}^2-u_{0,2})e^{-2u-2u_1}-10{A}\lambda_2(u_{0,1}^2+u_{0,2})\\&+5u_{0,1}\Big(4A^2\lambda_1^2(5e^{2u}-2e^{2u_1})e^{-4u-4u_1}+6{A}\lambda_1\lambda_2e^{-2u_{1}}+\lambda_2^2(e^{2u}-7e^{-2u_1})\Big)\\&+20A^2\lambda_1^3(e^{-4u-4u_1}+4e^{-2u-6u_1})-10{A}\lambda_1^2\lambda_2(e^{-4u}-3e^{-2u-2u_{1}}-10e^{-4u_1})\\&+5A^2\lambda_1\lambda_2^2(4e^{-2u_1}-7e^{-2u})-25{A}\lambda_2^3e^{2u_1}-C_1'Lf,\quad A=\sqrt{e^{2u_1}+e^{2u_0}}.}
Then we apply the operators  $T^{-1}\partial _{u_1}$ and $Y_{-1}$ to both sides of the obtained relation. In the first case we arrive at an equation of the form:
\eqa{\left(4\lambda_1^2e^{-4u}+4\lambda_1u_{0,1}e^{-2u}+\lambda_2{B}(u_{0,1}e^{2u}-3\lambda_1)+\lambda_2^2e^{2u}\right)\pfrac{^2}{u_{0,2}^2}g^{(3)}\\+(\lambda_2e^{2u}{B}-2\lambda_1e^{-2u})\pfrac{^2}{u_{0,2}\partial u_{0,1}}g^{(3)}+\pfrac{^2}{u_{0,2}\partial u}g^{(3)}=-4C_1\lambda_2{B}(3\lambda_1+u_{0,1}e^{2u})\\+4C_1\left(8\lambda_1^2e^{-4u}-\lambda_2^2e^{2u}+8\lambda_1u_{0,1}e^{-2u}\right)+40u_{0,1}(\lambda_2^2e^{2u}+7\lambda_1^2e^{-4u})\\+20\lambda_1e^{-2u}(u_{0,2}-2u_{0,1}^2)+30\lambda_1(\lambda_2^2-4\lambda_1^2e^{-6u})-10\lambda_2e^{2u}{B}(u_{0,2}+u_{0,1}^2)\\+15\lambda_2{B}(4\lambda_1^2e^{-2u}-\lambda_2^2e^{4u}+2\lambda_1u_{0,1})\\-C_1'\left(4\lambda_1e^{-2u}+\lambda_2{e^{2u}}B\right),\quad B=1/\sqrt{e^{2u}+e^{2u_{-1}}}.\label{g3_1}} The second one is the same up to the transformation $\lambda_1\to-\lambda_1,\lambda_2\to-\lambda_2,u_{-1}\to u_1$. We can compare in the equation \eqref{g3_1} coefficients in front of the independent variable $B$ and derive four equations. From the consistency condition of these equations we get $$C_1'(x)=0,\quad C_1(x)=C_1.$$ Summarizing the reasoning above we find the solution:
\eqa{g^{(3)}=&-2u_{0,2}(2\lambda_2^2e^{2u}+2\lambda_1^2e^{-4u}+2u_{0,1}^2-u_{0,2})C_1\\&-5u_{0,2}u_{0,1}(3\lambda_2^2e^{2u}-12\lambda_1^2e^{-4u}+u_{0,2})+C_3(x)u_{0,2}+g^{(4)}(x,u,u_{0,1}).}

In the next step we observe that equation \eqref{eq1_ts} is quadratic in $u_{0,2}$. Moreover, the second derivative with respect to $u_{0,2}$ reads as:
\seqs{-6C_1\lambda_2\sqrt{e^{2u}+e^{2u_{1,0}}}=0.} Hence we get $$C_1=0.$$ The first derivative with respect to $u_{0,2}$ coincides with:
\eqa{ (T-1)&\left(\pfrac{}{u_{0,1}}g^{(4)}-5u_{0,1}^4+270\lambda_1^2u_{0,1}^2e^{-4u}-5(\lambda_2^2e^{2u}+\lambda_1^2e^{-4u})^2+C_2'(u_{0,1}-3\lambda_1e^{-2u})\right.\\&+\left.\frac12C_2(3u_{0,1}^2+18\lambda_1u_{0,1}e^{-2u}+3\lambda_2^2e^{2u}+3\lambda_1^2e^{-4u})\right)+6\lambda_1e^{-2u}(3u_{0,1}C_2-C_2')=0.\label{g4_1}} 
Obviously equation \eqref{g4_1} implies 
$$C_2(x)=0,$$
\eq{g^{(4)}=u_{0,1}^5-90\lambda_1^2u_{0,1}^3e^{-4u}+5(\lambda_2^2e^{2u}+\lambda_1^2e^{-4u})^2u_{0,1}+C_4(x)u_{0,1}+g^{(5)}(x,u).}

Now equation \eqref{eq1_ts} has the quadratic term in $u_{0,1}$ with the coefficient $C_3(x)$, so we should take $$C_3(x)=0.$$The remaining part in \eqref{eq1_ts} has the form:
\eq{(T-1)\left(u_{0,1}\pfrac{}{u_0}g^{(5)}+\pfrac{}{x}g^{(5)}+C_4'u_{0,1}\right)=\pfrac{f}{u_{0}}g^{(5)}+\pfrac{f}{u_{1}}Tg^{(5)}.\label{eq5_1}}
Let us apply the operator $\pfrac{^2}{u_{0,1}\partial u_0}$ to \eqref{eq5_1} and find $$\pfrac{^2}{u_0^2}g^{(5)}=0,\quad g^{(5)}=C_5(x)u_0+C_6(x).$$

Now equation \eqref{eq5_1} does not depend on $u_{0,1}$. This equation must be fulfilled identically for all values of $u_0$ and $u_1$, in particular, for $u_{1}=u_0$ we get:
$$\sqrt 2\lambda_2(C_4'+C_5-C_6)e^u+2\lambda_1(C_4'+C_5+2C_6)e^{-2u}+4C_5\lambda_1 u e^{-2u}-\sqrt 2u\lambda_2 C_5e^{-u}=0.$$ Functions $e^u,e^{-2u},ue^u,ue^{-2u}$ are linearly independent, so we get $$C_5=0,\quad C_4'-C_6=0,\quad C_4'+2C_6=0.$$  We find $$C_6=0,\quad C_4(x)=c=\hbox{const}.$$
Let us take $c=0$ since this term is responsible for the classical symmetry. Finally, we get the generalized symmetry \eqref{u_t}.


\bigskip

We proceed to generalized symmetries in the direction of $n$. It is easily proved that there is no generalized symmetry of the form \eqref{volt} with $M=1$, therefore the simplest symmetry of this kind should be of the form
\eq{\partial_\tau u_{0}=h(u_{-2},u_{-1},u,u_{1},u_2).\label{inb}}
Upon closer examination of the equation \eqref{eq1_}, adapted to our case, we can see that it is actually a linear expression with respect to the variable $u_{0,1}$. Therefore the coefficient in front of $u_{0,1}$ vanishes. This condition yields the following equation:
\eq{(T-1)\sum_{k=-2}^2\pfrac{}{u_k} h(u_{-2},u_{-1},u,u_{1},u_2)=0}
that obviously implies a linear first order PDE for $h$
\seq{\sum_{k=-2}^2\pfrac{}{u_k} h(u_{-2},u_{-1},u,u_{1},u_2)=C_0,} where $C_0$ is an arbitrary constant. General solution to the latter equation is easily found 
\eq{h(u_{-2},u_{-1},u,u_{1},u_2)=h^{(1)}(u_{-2}-u_{-1},u_{-1}-u,u-u_1,u_1-u_2)+C_0u_0.}
Let us rewrite the original lattice \eqref{tz} in terms of the variable $y_n=u_n-u_{n+1}$:
\eq{\dfrac{y}{x}=\dfrac{(u_{0}-u_1)}{x}=\lambda_1e^{-2u}(1+e^{2y})+\lambda_2e^{u-y}\sqrt{1+e^{2y}}}
In order to bring the lattice to a rational form  we make in addition a point transformation from $y$ to $v$ defined by formulas $$v=\sqrt{1+e^{2y}}+e^y,\quad e^y=\frac{v^2-1}{2v},\quad \sqrt{1+e^{2y}}=\frac{1+v^2}{2v}.$$ 
In terms of the variable $v_k$ we have:
\eqa{\dfrac{v_k}{x}=\lambda_1V_ke^{-2u}+\lambda_2U_ke^u,} where $V_k$ and $U_k$ are functions depending on a finite set of the shifts $\{v_n\}$ only, and can be found recursively:
\seqs{V_0=-\frac{v_0^4-1}{4v_0},\quad U_0=-v_0,\\ V_k=\frac{(v^2-1)^2}{4v^2}TV_{k-1},\quad U_k=\frac{2v}{v^2-1}TU_{k-1},\ k\in\N,\\ V_{-k}=T^{-1}\frac{4v}{(v^2-1)^2}V_{-k+1},\quad U_{-k}=T^{-1}\frac{v^2-1}{2v}U_{-k+1},\ k\in\N.}

We look for the right hand side $h$ of the symmetry \eqref{inb} in the form
\eq{h(u_{-2},u_{-1},u,u_{1},u_2)=H(v_{-2},v_{-1},v,v_1)+C_0u_0.} Note that now the set of functions $u:=u_0,v_0,v_{1},v_{-1},\ldots$ defines a new set of dynamical variables. By passing in the equation \eqref{eq1_} to this set of the variables we arrive at:
\eq{\lambda_2e^uM_1+\lambda_1e^{-2u}M_2+C_0u(v^2+1)\left(\frac{\lambda_1(v^2+1)}{2v^2}e^{-2u}-\frac{\lambda_2}{v^2-1}e^u\right)=0,\label{def}}where
\eqa{M_1=\sum_{k=-2}^{+2}U_{k}\pfrac{}{v_{k}}(T-1)H+\frac{v^2-1}{v^2+1}H+\frac{4v^2}{v^4-1}TH\\-C_0\left(U_0\frac{v^2+1}{v(v^2-1)}+\frac{4v^2}{v^4-1}\log\frac{2v}{v^2-1}\right),\\
M_2=\sum_{k=-2}^{+2}V_{k}\pfrac{}{v_{k}}(T-1)H-2H-\frac{v^4-1}{2v^2}TH\\-C_0\left(V_0\frac{v^2+1}{v(v^2-1)}-\frac{(v^2-1)^2}{2v^2}\log\frac{2v}{v^2-1}\right).\nonumber}
Since functions $e^u,e^{-2u},ue^u, ue^{-2u}$ are linearly independent we immediately obtain from \eqref{def}:
$$C_0=0,\quad M_1=0,\quad M_2=0.$$ 
The last two equations are functional differential equations for function $H(v_{-2},v_{-1},v,v_1)$. Dependence of $H$ on the variables $v_{-2},v_{1}$ can be specified due to the system of partial differential equations\eqa{T^{-1}\pfrac{}{v_2}M_1=0,\quad T^{-1}\pfrac{}{v_2}M_2=0,\\
\pfrac{}{v_{-2}}M_1=0,\quad \pfrac{}{v_{-2}}M_2=0.}Solution of this system is given by:
\eqa{H=&\frac{C_1v(v_{-1}^2+1)}{\left(v_1^2\frac{(v+1)^2}{v-1}^2+1\right)(v^2(v_{-1}+1)^2+(v_{-1}-1)^2)}\\&+\frac{C_2(v^2+1)v_{-1}^2(v_{-1}^2-1)}{(v^2(v_{-1}+1)^2+(v_{-1}-1)^2)\left(v_{-1}^2+\frac{(v_{-2}-1)^2}{(v_{-2}+1)^2}\right)}+H^{(1)}(v_{-1},v).}
Any of the equations 
$$\pfrac{^2}{v_{-1}\partial v_1}M_1=0,\quad \pfrac{^2}{v_{-1}\partial v_1}M_2=0$$ implies $C_2=C_1/2.$
Similarly from the system 
$$\pfrac{}{v_{-1}}M_1=0,\quad \pfrac{}{v_{-1}}M_2=0,\quad T^{-1}\pfrac{}{v_{1}}M_1=0,\quad T^{-1}\pfrac{}{v_{1}}M_2=0$$ one can find
$$H^{(1)}=-\frac{C_1}2\frac{(v_{-1}+1)v^2-(v_{-1}^2+1)v-v_{-1}+1}{v(v_{-1}+1)^2+(v_{-1}-1)^2}+C_3.$$
And now from $M_1=0$ or from $M_2=0$ we find $C_3=-C_1/4.$ For simplicity we put $C_1=4$ and finally get \eqref{u_tau}.

\subsection{Equations in variable $v_n$}
We can easily rewrite generalized symmetry \eqref{u_tau} in terms of the variable $v$:
\eqa{\partial_\tau v=v(v^2-1)&\left(v^2+1-\frac{4v_{-1}^2}{v^2+1}T^{-1}-\frac{(v_1^2-1)^2}{v^2+1}T\right)\\&\frac{(v_{1}^2+1)(v_{-1}^2+1)}{(v^2(v_{-1}+1)^2+(v_{-1}-1)^2)(v_1(v+1)^2+(v-1)^2)}\label{v_tau}}
We can rewrite the semi-discrete equation \eqref{tz} in terms of $v$ as well. However to this end we have to solve a cubic equation:
\eq{\lambda_2v p^3+v_{0,x}p^2+\lambda_1\frac{v^2-1}{4v}=0.} If $p(v_0,v_{0,x})$ is a solution of the cubic equation, then $$p(v_0,v_{0,x})\frac{2v}{1-v^2}=p(v_1,v_{1,x})$$ is semi-discrete equation, defining the Backlund transformation for the equation \eqref{v_tau}.

\subsection{ Continuum limit}
Here we compute the  continuum limit of the semi-discrete Tzitzeika equation \eqref{tz} and of its generalized symmetries. 
To do this we apply the transformation:
\eq{u_n(x,t,\tau)=U\left(x,y-\frac49\varepsilon \tau,t,\frac{2\varepsilon^5}{135} \tau\right),\quad \lambda_1=\varepsilon a/2,\quad\lambda_2=\varepsilon b/2.\label{lim}}
Then, as $\varepsilon\to 0$ from \eqref{tz} we get:
\eq{U_{x,y}=ae^{-2U}+be^U\label{tz_xy}.}
The same substitution \eqref{lim} into \eqref{u_t} and \eqref{u_tau} gives rise for $\varepsilon\to 0$ to equations \eqs{\partial_t U&=&U_{xxxxx}+5U_{xxx}(U_{xx}-U_{x}^2)-5U_{xx}^2U_{x}+U_{x}^5,\label{91}\\ \partial_{\theta} U&=&U_{yyyyy}+5U_{yyy}(U_{yy}-U_{y}^2)-5U_{yy}^2U_{y}+U_{y}^5,\quad \theta=\frac{2\varepsilon^5}{135} \tau.\nonumber} 
The obtained equations are well known $x-$ and $y-$symmetries of the Tzitzeika equation \eqref{tz_xy}. 

\section*{Conclusions} The article proposes a method for finding generalized symmetries for semi-discrete lattices of hyperbolic type. We hope that the method will find fruitful application in studying the problem of an integrable classification of lattices of this type. Despite the fact that a large number of integrable semi-discrete models of hyperbolic type have already been obtained within the framework of various discretization algorithms, including the Backlund transformations, this class requires further study: in our opinion, it may contain new integrable models. Our hypothesis is supported by the example of the Tzizeica equation given in this article. 

\paragraph{Acknowledgments.} The work of GRN is partially supported in the framework of executing the development
program of Scientific Educational Mathematical Center of Privolzhsky Federal Area, additional agreement no. 075-02-2020-1421/1 to agreement no. 075-02-2020-1421.


\begin{thebibliography}{20}
\bibitem{Adler} V.E. Adler,  \emph{On a discrete analog of the Tzitzeica equation}, (2011) arXiv preprint arXiv:1103.5139.
\bibitem{b22} L. Bianchi, \emph{Lezioni di geometria differenziale}: in 2 vol. Pisa: Spoerri, 1922.
\bibitem{Dodd}  R.K. Dodd, R.K. Bullough, \emph{Polynomial conserved densities for the sine-Gordon equation}. Proc. Roy. Soc. London A {\bf 352} (1977) 481--503.

\bibitem{GarifullinHabibullin} R.N. Garifullin, E.V. Gudkova, I.T. Habibullin,  \emph{Method for searching higher symmetries for quad-graph equations}. J. Phys. A {\bf 44} (2011), no. 32, 325202, 16 pp.
\bibitem{is79} N. H. Ibragimov, A. B. Shabat, \emph{Korteweg–de Vries equation from the group standpoint}, Dokl. Akad. Nauk SSSR, 244:1 (1979), 57-61.
\bibitem{is80}N. Kh. Ibragimov, A. B. Shabat, \emph{Evolutionary equations with nontrivial Lie–B\"acklund group}, Funktsional. Anal. i Prilozhen., 14:1 (1980), 25-36; English transl. in Funct. Anal. Appl., 14:1 (1980), 19-28.

\bibitem{Habibullin2008} I. Habibullin, N. Zheltukhina, A. Pekcan, \emph{On the classification of Darboux integrable chains}. J. Math. Phys. {\bf 49} (2008), no. 10, 102702, 39 pp.
\bibitem{Habibullin2009} I. Habibullin, N. Zheltukhina, A. Pekcan,  \emph{Complete list of Darboux integrable chains of the form $t_{1x}=t_x+d(t,t_1)$}. J. Math. Phys. {\bf 50} (2009), no. 10, 102710, 23 pp.




\bibitem{h05}I.T. Habibullin, \emph{Characteristic algebras of fully discrete hyperbolic type equations}, SIGMA  Symmetry Integrability Geom. Methods Appl. {\bf 1} (2005) 023.

\bibitem{ly09} D. Levi and R.I. Yamilov, \emph{The generalized symmetry method for discrete equations}, J. Phys. A: Math. Theor. {\bf 42} (2009) 454012 (18pp).

\bibitem{ly11}D. Levi and R.I. Yamilov, \emph{Generalized symmetry integrability test for discrete equations on the square lattice}, J. Phys. A: Math. Theor. {\bf 44} (2011) 145207 (22pp).
\bibitem{ms12} A.G. Meshkov, V.V. Sokolov. {\it Integrable evolution equations with a constant separant} // Ufimsk. Mat. Zh. {\bf 4}:3. 104--154 (2012) [Engl. trans.: Ufa Math. Journal {\bf 4}:3. 104--152 (2012).]
\bibitem{Mikh}  A.V. Mikhailov,  \emph{Integrability of a two-dimensional generalization of the Toda chain}. Soviet Phys. JETP Lett. {\bf 30} (1979) 414--418.

\bibitem{mwx11} A. Mikhailov, J.P. Wang and P. Xenitidis, \emph{Recursion operators, conservation laws, and integrability conditions for difference equations}, Teoret. Mat. Fiz. {\bf 167}:1 (2011) 23--49 (in Russian); English transl. in Theor. Math. Phys.  {\bf 167}:1 (2011) 421--443.
\bibitem{mx13}A.V. Mikhailov  and P. Xenitidis, \emph{Second order integrability conditions for difference equations: an integrable equation}, Lett.  Math. Phys. {\bf 104}:4  (2014)  431--450.


\bibitem{msy87}A. V. Mikhailov, A. B. Shabat, R. I. Yamilov, \emph{The symmetry approach to the classification of non-linear equations. Complete lists of integrable systems}, Uspekhi Mat. Nauk, 42:4(256) (1987), 3-53; English transl. in Russian Math. Surveys, 42:4 (1987), 1-63 

\bibitem{shl}C. Scimiterna, M. Hay and  D. Levi, {\it On the integrability of a new lattice equation found by multiple scale analysis},  J. Phys. A: Math. Theor. {\bf 47} (2014) 265204 (16 pp), arXiv:1401.5691.
\bibitem{ss82} S.I. Svinolupov, V.V. Sokolov, \emph{Evolution equations with nontrivial conservative laws}. Funct Anal Its Appl {\bf 16}, 317-319 (1982)

\bibitem{Tzitz}  G. Tzitzeica, \emph{Sur une nouvelle classe de surfaces}. Rendiconti del Circolo Matematico di Palermo {\bf 25}:1 (1907) 180--187.
\bibitem{vs93}A.P. Veselov, A.B. Shabat, \emph{Dressing Chains and Spectral Theory of the Schr\"odinger Operator}, Funct. Anal. Appl., {\bf 27}:2 (1993), 81-96.
\bibitem{Xenitidis} P. Xenitidis, \emph{Determining the symmetries of difference equations}. Proceedings of the Royal Society A, {\bf 474}(2219), (2018) 20180340.
\bibitem{y83}R.I. Yamilov, {\it Classification of discrete evolution equations}, Uspekhi Mat. Nauk {\bf 38}:6 (1983) 155--156 [in Russian].
\bibitem{Yamilov1990} R.I. Yamilov,   \emph{Invertible changes of variables generated by Backlund transformations}. Theoretical and Mathematical Physics, {\bf 85}(3), (1990) 1269-1275.
\bibitem{Yamilov06} R. Yamilov,  \emph{Symmetries as integrability criteria for differential difference equations}. Journal of Physics A: Mathematical and General, {\bf 39}(45), (2006) R541.
\bibitem{Shabat}  A.V. Zhiber, A.B. Shabat, \emph{Klein-Gordon equations with a nontrivial group}. Soviet Phys. Doklady {\bf 24} (1979) 607.

 \end{thebibliography}
\end{document}